\documentclass[12pt]{article}
\usepackage[pdftex]{graphicx}
\usepackage[english]{babel}
\usepackage[utf8]{inputenc}
\usepackage{mathptmx} 

\usepackage{hyperref}

\begin{document}

\title{Observation of the prior earthquake effect on the flux of environmental neutrons, gamma-radiation, and on the local electric field in Tien~Shan mountain}

\author{
N.~M.~Salikhov,$^1$
A.~L.~Shepetov,$^{2,3}$
A.~P.~Chubenko,$^2$ \\
O.~N.~Kryakunova,$^1$
G.~D.~Pak$^1$
}

\hypersetup{
  pdftitle =
    {Earthquake effect on the environmental neutrons, gamma radiation, and on the local electric field},
  pdfauthor =
    { N.~M.~Salikhov, A.~L.~Shepetov, A.~P.~Chubenko, O.~N.~Kryakunova, G.~D.~Pak}
}

\maketitle

\noindent
$^{1}${Ionosphere Institute, Almaty, Kazakhstan}\\
$^{2}${P.~N.~Lebedev Physics Institute, Moscow, Russia}\\
$^{3}${Tien~Shan Mountain Cosmic Ray Station, Almaty, Kazakhstan}

\section*{Abstract}
A search for the possible precursors of an earthquake and its effect on the data series of diverse geophysical parameters has been made in the mountain of Northern Tien~Shan. The complex installation included the NM64~type neutron supermonitor, detectors of the environmental low-energy neutrons, the scintillation gamma-detector, the sensor of the local electric field, a seismograph, and a weather-station. The specialized data filtration methodic was elaborated with an enhanced sensitivity to the transient signals of seismic origin. On the eve of, and after a 5.4~magnitude earthquake the fine features in temporal behavior of the intensity of low-energy neutron and gamma-radiation background, so as irregularities of the local electric field were observed which seem to be in a close correlation with each other. These results may be an evidence of the possibility of experimental identification of earthquake's precursors in the time up to $2-3$~days before the beginning of a period of intensive tectonic activity.

\section{Introduction}

The urgency of the short-time forecasting of earthquakes, especially of high-magnitude ones, and the search for their possible precursors is caused by the huge social-economic losses resulting from disastrous consequences of these natural phenomena. In last time, much attention is paid to investigation of the mutual correlations between a number of different geophysical, atmospheric, and astrophysical characteristics which may be used for diagnostics of the internal centers of seismic tension from the viewpoint of a timely earthquake forecast.

One of the possible effects being, in particular, under consideration at present time as a probe candidate of inner layers of the Earth's crust is the intensity variation of the near-Earth environmental flux of slow neutrons. It is generally believed that at least a part of this flux, together with atmospheric neutrons originating in interaction of energetic cosmic ray particles, constitute the neutrons being born through the $(\alpha,n)$~type reactions of the crust's nuclei with $\alpha$-particles which arise in decays of the natural radioactive elements, primarily radon $^{222}Rn$ and thoron $^{220}Rn$. After their thermalization in multiple scattering with nuclei, slow neutrons of terrestrial origin can reach the surface of the Earth and be detected afterwards \cite{seismoneutro_mgu0}. Thus, a perceptible correlation has been observed between the local intensity of low-energy neutrons and the Moon phase, which, accordingly to the most appropriate interpretation, may be connected with a subsidiary effusion of the $\alpha$-radioactive gases in additional stress of the inner crust during the new- and full moon periods under the summed gravitational influence of the Moon and Sun  \cite{seismoneutro_mgu1,seismoneutro_baksan}. Similar effect on the inner layers may be caused by accumulated mechanical strain in a seismically active region which disturbances, in turn, should be accompanied with corresponding deviation of the environmental neutron background from its equilibrium state. Indeed, significant perturbations of the registered intensity of neutron counting rate have been repeatedly reported in the period of seismic activity, both in local scale \cite{seismoneutro_baksan,seismoneutro_almaty}, and globally \cite{seismoneutro_mgu2}.

As for the atmospheric neutron component of the cosmic ray origin, in the range of low energies its intensity is governed, in particular, by the current state of interplanetary medium. Together with direct influence on the intensity of high-energy cosmic ray particles --- predecessors of the slow environmental neutrons, any abrupt changes in these conditions, like that being observed after a solar flare, may trigger the processes of seismic activity inside the stressed regions of the Earth's crust which circumstance stipulates another, indirect connection mechanism between the earthquakes and neutron variations \cite{seismoneutro_mgu2}

Another geophysical phenomenon which seems to be interesting from the viewpoint of earthquake forecasting is the behavior of atmospheric electricity in preceding period. As it has been stated in \cite{seismoele_morgunov2,seismoele_morgunov} electromagnetic emissions in a frequency range from kHz to MHz may be produced by opening the cracks in creep processes which develop in the mechanically loaded areas of the Earth's crust before an earthquake. Possibility to use  this phenomenon for prediction of the spatial location of a nearest earthquake's epicenter is illustrated by the works \cite{seismoele_greece1,seismoele_greece2,seismoele_greece3, seismoele_china}.

The influence of the low-frequency electromagnetic fields on the region of iono\-sphe\-re–-mag\-ne\-to\-sphe\-re transition may cause anomalous precipitations of the charged particles from the inner radiation belt which have been many times registered in a earthquake preceding period from the low-orbit satellites \cite{seismocosmo1,seismocosmo2,seismocosmo3}. One could expect that the effects of this kind may cause also deviations in the usual behavior of different components of the secondary cosmic rays deep in the atmosphere, in particular, in the intensity of the electron- and gamma-radiation background.

Quite a different approach to the problem of the short-time earthquake forecast is suggested in \cite{seismocorri_kamcha2}. Because of the interrelated and complex nature of seismic events it is advisable to search for specific features of correlation pattern between a number of heterogeneous phenomenological parameters (the ground water salinity and the amplitude of geoacoustic emissions in case of \cite{seismocorri_kamcha2}) which may signal the nearness of an earthquake. The necessity to fix simultaneously a wide range of geophysical characteristics for the purpose stipulates a preferable using of the complex experimental installations which is just the case of present paper.

On the date of 1~May, 2011, a 5.4~magnitude earthquake has occurred in the vicinity of Almaty city in Kazakhstan Republic. Accordingly to the data of European-Medi\-ter\-ra\-ne\-an Seismic Center, its precise moment is 02:31:29~UT while an aftershock series lasted over the next ten days. The effects of this earthquake on a number of geophysical parameters were registered by detectors of the Tien~Shan Mountain Cosmic Ray Station's complex placed 95.6~km apart from epicenter. In particular, this installation includes a number of neutron detectors, both for the hadronic component of the high energy cosmic rays (the NM64~type neutron supermonitor), and for the low-energy neutron background from the nearest environment (detectors of thermal neutrons consisting of a set of unshielded proportional neutron counters); the gamma-detector on the basis of a NaI~scintillation crystal which is capable to measure the radiation intensity in the range of  $20-1200$~keV, separately in six energy diapasons; the electrometer for registration of the induced charge and the ionic conductivity of atmosphere; the high-sensitive detector of the low-frequency (infrasonic) variations of atmospheric pressure, and the standard common purpose weather station for registration of meteorological parameters (air pressure, temperature, humidity etc). In detail, the system of neutron detectors of the Tien~Shan mountain station is described in \cite{jopg2008} and \cite{thunderour2011b}; the system of gamma-radiation scintillation detectors --- in \cite{thunderour2009}
and \cite{thunderour2011}. The data registration procedure in considered measurements consisted of a continuous monitoring of detector signals whereas the intensities of neutron pulses, those of gamma-scintillations, and the weather data were stored with a 1-min temporal resolution, and the measurements of the electric field --- with a resolution of one second.

The data obtained in this seismic event are discussed in the present paper.

\section{Experimental Results and Discussion}

\begin{figure}
\begin{center}
\includegraphics[width=0.95\textwidth,trim=0mm 15mm 0mm 0mm]{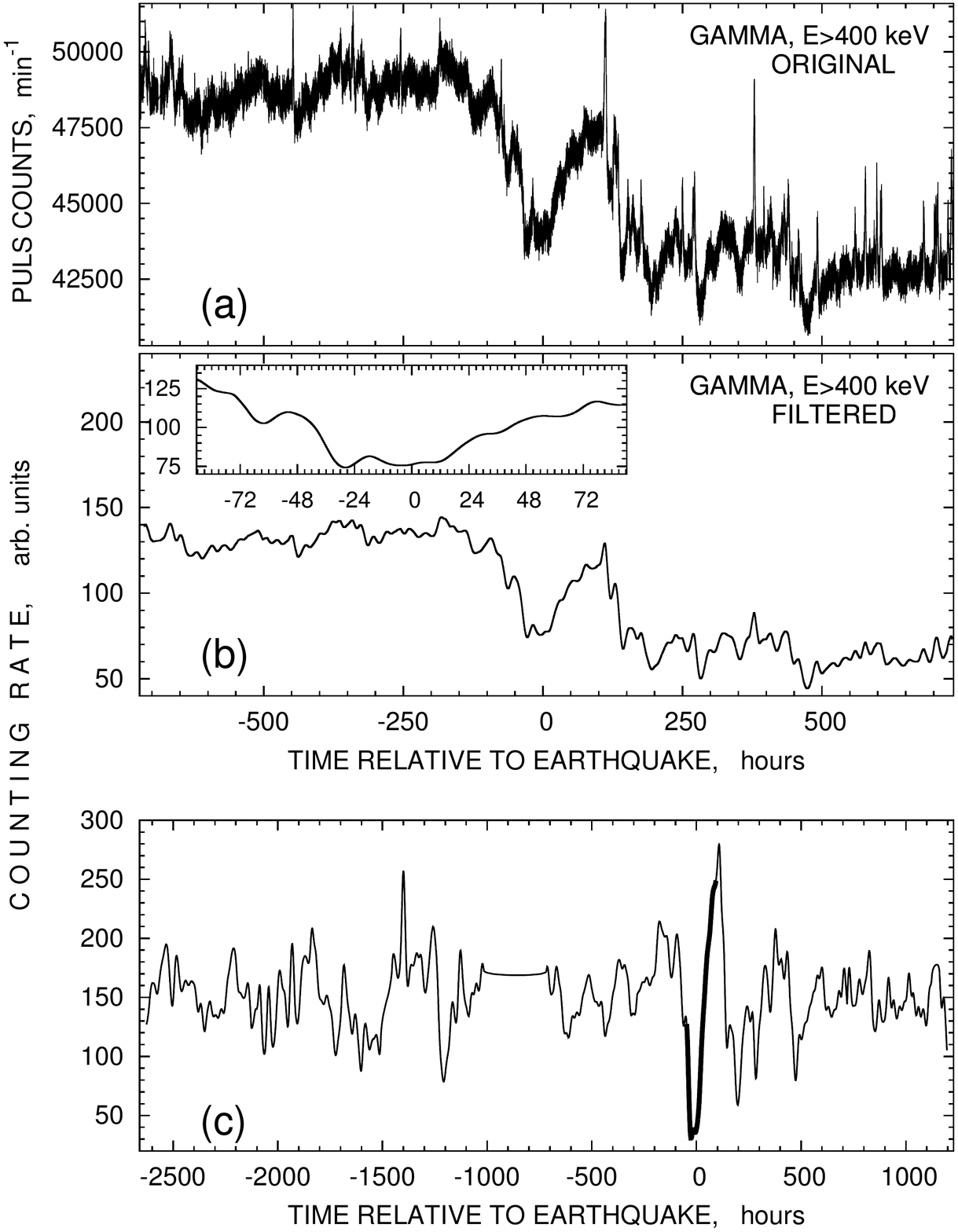}
\end{center}
\caption{Experimental data: the 1-min resolution time history of the intensity of $E_{\gamma}\ge400$~keV gamma-radiation in its original form (a); same distribution after being smoothed with a 500~min low-pass filter (b); the continuous long-term record of gamma-radiation intensity over the whole 0.5~year long period after smoothing and subtracting its seasonal trend (c). Each time, zero point of abscissa axis coincides with the 1~May~2011~earthquake moment.}
\label{fig1}
\end{figure}

{\it Gamma-radiation.}
Time history of the intensity of gamma-radiation registered on the  stretch of $\pm$one month around the earthquake is shown in its original form in the upper plot (a) of the figure~\ref{fig1}. The lower energy threshold of detected gamma-quanta in these measurements was 400~keV, and the time axis of the plot is centered relatively to the earthquake of 1~May, 2011, its zero point coinciding strictly with the earthquake moment. To eliminate accidental interferences of electromagnetic nature which sometimes occurred irregularly during the measurements the initial experimental data are presented in this plot after being piped through the moving average filter.

As it is seen in the upper panel of figure~\ref{fig1}, the record of gamma-radiation intensity demonstrates two kinds of irregularity. First, there are numerous transitory enhancements with typical duration of the order of an hour which are met in multitude on considered plot. This is a common effect which usually accompanies precipitation, in particularly rains and hails, and has been repeatedly observed before, both in the Tien~Shan experiment \cite{rainsour2009} and in the measurements of other groups (e.g., \cite{rains2,rains3}). Against the end of presented period (second half of May 2011), these enhancements occur more frequently, in connection with the beginning of the summer rain time in the mountain of Northern Tien~Shan.

More interesting is another irregularity, a noticeable dip in gamma-radiation intensity seen around zero point of the time axis, i.e. in the immediate vicinity of the 1~May, 2011 earthquake. To stress this effect more prominently and to get rid of the short-time fluctuations, the experimentally measured intensity distribution was additionally re-calculated using a digital low-pass filter with the time constant $\tau=500$~min, and the result of this procedure is presented in the middle (b) plot of the figure~\ref{fig1}. It is seen both on the plots (a) and (b) that, starting since 29~April, 1.5 days before the earthquake, the intensity of gamma-radiation signals has diminished constantly and reached its minimum on 30~April, immediately on the eve of the primary seismic shock, while after this shock happened an abrupt growth in the counting rate.  It should be noted also, that, accordingly to accompanying data of the Tien~Shan weather station, abundant precipitations have taken place, and the relative humidity remained about 100\% during the whole day of 30~April, while on 1--4~May there was a period of the fair cloudless weather, so the variation of gamma-ray intensity during the days of 29~April--1~May occurred in an opposite direction relatively to its usual behavior in rainy days: instead to grow up it drops at a rain and raises in a clear time.

The bottom (c) plot of the figure~\ref{fig1} demonstrates the continuous record of temporal behavior of the intensity of gamma-radiation as it has been regularly registered at Tien~Shan station over a larger time stretch (of about 0.5~year, since 1~January~2011 and up to the middle of June~2011). As before, the initial pulse counts of gamma-detector presented here are smoothed with a low-pass filter, and, additionally, the seasonal long-term trend in the data which is generally caused by  temperature drift of scintillator's photomultiplier tube is subtracted from this distribution. As it is seen on the bottom plot, after the filtering procedure being applied the only characteristically sharp irregularity in gamma-radiation intensity over the whole presented period remains in the vicinity of the earthquake moment (which is put in bold in the figure~\ref{fig1}(c)). This uniqueness is an evidence against any random coincidence between both rare events.

\begin{figure}
\begin{center}
\includegraphics[width=0.91\textwidth,trim=0mm 140mm 0mm 0mm]{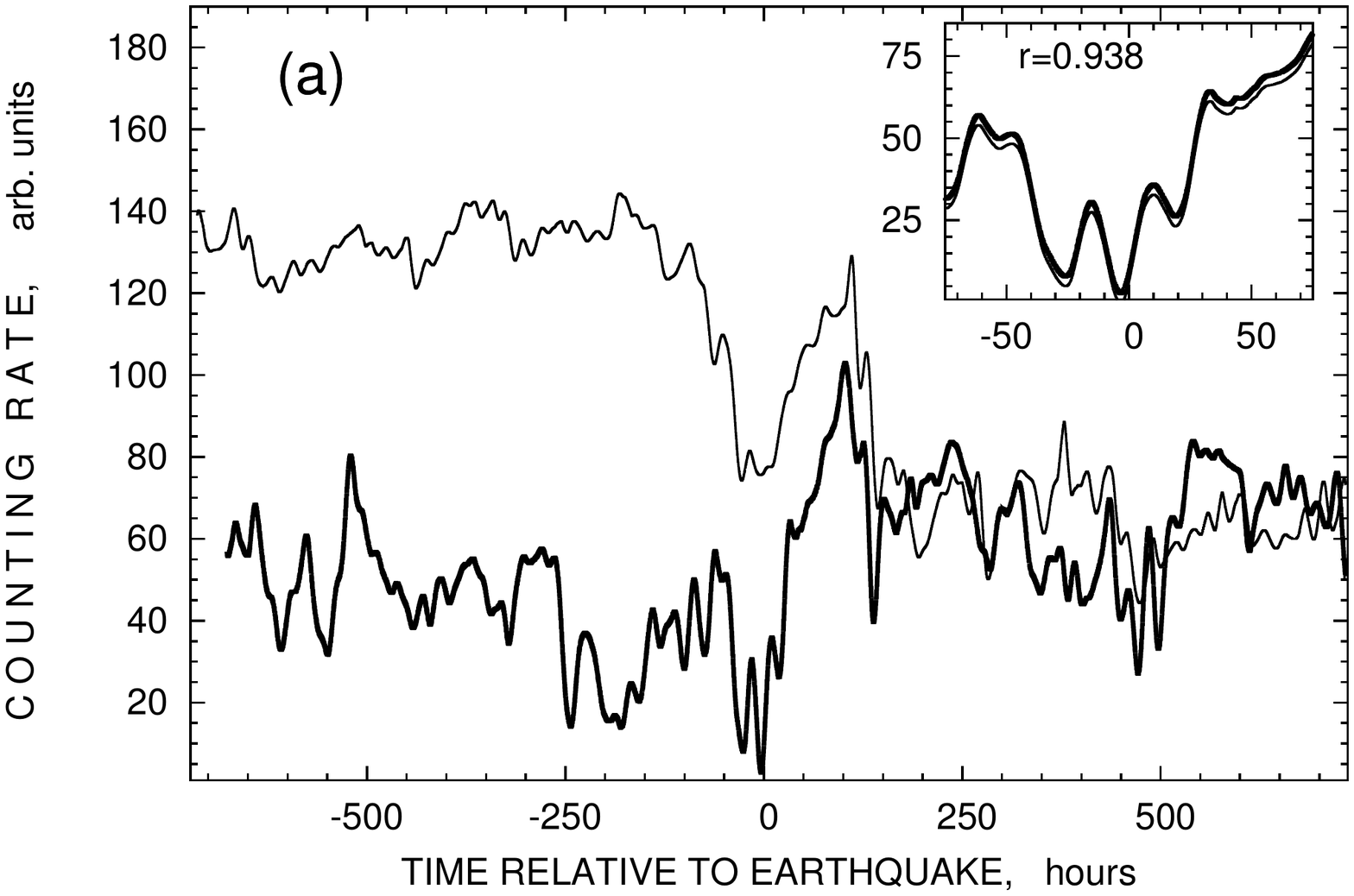}
\includegraphics[width=0.95\textwidth,clip,trim=15mm 10mm 15mm 15mm]{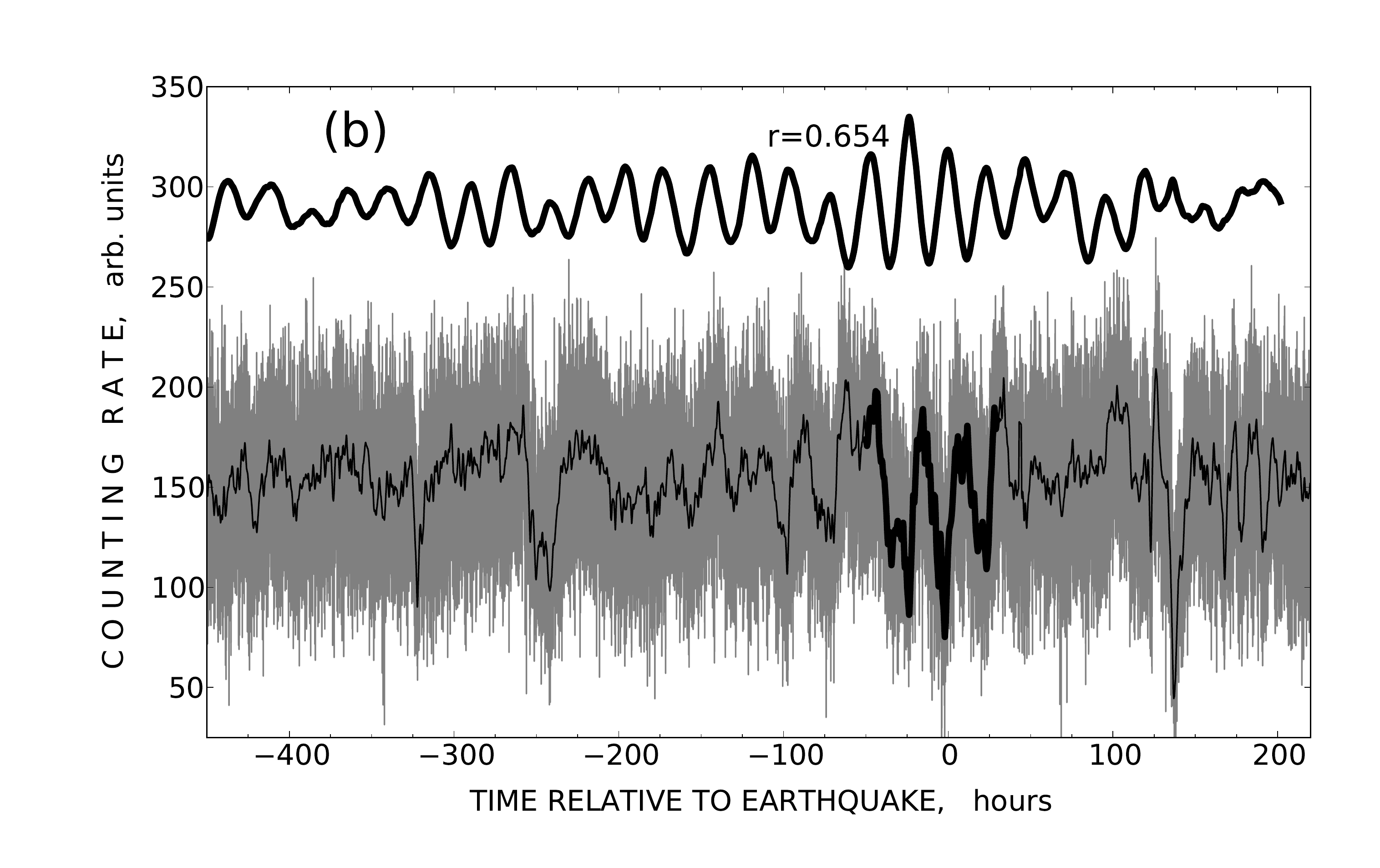}
\end{center}
\caption{Upper panel: time history of the counting rate of thermal neutrons around the 1~May~2011~earthquake moment (bold line) shown in comparison with the gamma-radiation intensity (thin line); lower panel: diurnal rhythmic and auto-correlation function of neutron signal in the vicinity of earthquake (see text).}
\label{fig2}
\end{figure}

{\it Neutron signal.}
Temporal distribution of the intensity of environmental thermal neutron background obtained during the time both preceding and succeeding the earthquake date is shown, with a bold curve, in the upper panel (a) of the figure~\ref{fig2}. For comparison, the smoothed gamma-radiation distribution which has been spoken above is plotted on this panel as well with a thin curve. The low-energy neutron detector of the Tien~Shan station used in these measurements registered most effectively the neutrons in the range of thermal energies, $E_n \sim 10^{-2}-10^{-1}$~eV. Initial measurements of the neutron counting rate which have been held with a 1-min periodicity were corrected for variation of the atmospheric pressure~$p$ accordingly to the formula generally used in the cosmic ray variations study
\cite{carmichel_meteocoeff}: $i_{corr}=i\cdot exp(0.0072\cdot(p-p_0))$, where $i_{corr}$~and~$i$ are the corrected and original values of neutron signal intensity, and $p_0=675$~mbar is the mean atmospheric pressure at the altitude of Tien~Shan station. Similarly to the case of the gamma-radiation signal, the low-pass filtering procedure was applied to these data after the pressure correction.

It is seen in the figure~\ref{fig2}(a) that during the $40-50$~h long period before the 1~May~2011~earthquake the intensity of low-energy neutrons diminishes which behavior generally repeats the considered above behavior of gamma-radiation. (However, any irregularities in this period are totally absent in the signal of the high-energy ($E_n \ge 1$~GeV) neutrons and cosmic ray hadrons which have been registered simultaneously by the Tien~Shan neutron supermonitor). The mutual likeliness of the neutron and gamma-radiation signals immediately before the earthquake moment is illustrated by the incut to figure~\ref{fig2}(a) which demonstrates a quite high consistency of both distributions with correlation coefficient $r$ between them being about~0.94.

In contrast to the low-frequency time series of the upper panel in figure~\ref{fig2}, the distribution plotted on the figure~\ref{fig2}(b) with a thin gray line was obtained from original measurements of the neutron counting rates with application of a digital high pass filter calculated over 3000~succeeding data points (i.e. having the time constant of 3000~min). This procedure permits to eliminate any slow trend from and stress the fast variations in neutron signal. Averaging of this high-frequency distribution over 50~neighboring points leads to another envelope curve, shown with a black line in figure~\ref{fig2}(b) which is imposed on the gray one. As it is seen, the latter curve generally shows an obvious diurnal oscillation of neutron intensity; and mostly prominently this oscillation reveals itself just in the days which precede and succeed the earthquake moment (which is stressed with a bold portion of envelope line in the figure~\ref{fig2}(b)). If to take this 3~days long segment (the dates of 30~April, 1, and 2~May~2011) and to use it for calculation of an auto-correlation function with the same envelope curve then the maximum correlation coefficient, of the order of~0.65, is achieved, again, in the vicinity of the earthquake (see the bold sinusoid-like curve above in the plot of figure~2(b)). This analysis shows that some characteristic amplification of the diurnal rhythmic of environmental neutron signal has happened in the period of seismic activity.

\begin{figure}
\begin{center}
\includegraphics[width=\textwidth,trim=0mm 30mm 5mm 0mm]{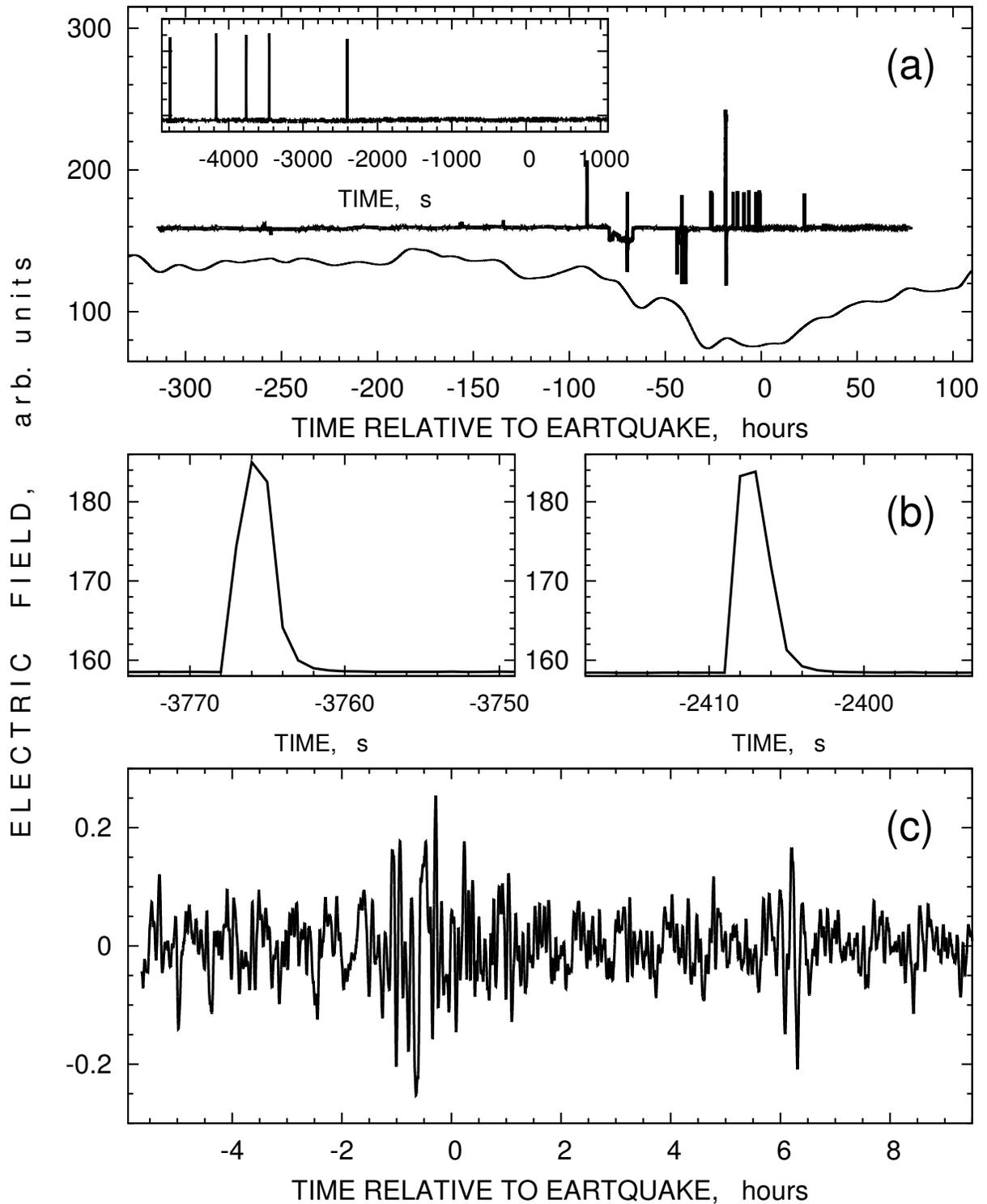}
\end{center}
\caption{(a): record of the strength of electric field in the time preceding the 1~May~2011 earthquake (thick line) in comparison with the gamma-radiation intensity (thin line); (b): typical structure of the short-time field irregularities on the example of two single spikes; (c): fine structure of the record of electric field after elimination of the spike-like outbursts and applying a low-pass filter.
Zero point of the time axis always coincides with the moment of the main earthquake shock.}
\label{fig3}
\end{figure}

{\it Electric field.}
Another anomaly observed during the days preceding to earthquake was found in the behavior of the local electric field which has been measured with a 1~s time periodicity by an electrometer facility placed 2.9~km apart from the detector complex of Tien~Shan station, and 91.7~km apart from epicenter. As it is seen in the upper panel of figure~\ref{fig3}, the unusual unidirectional spike-like outbursts of positive polarity appear some days before the earthquake in the record of electric field, simultaneously with the drop of gamma-radiation intensity (shown again with a thin curve on the same figure), their frequency growing constantly until the moment of the first shock, and disappearing completely at the end of aftershock period.

The powerful separate enhancements in the strength of electric field, each $3-5$~s in duration (see the plots (b) in figure~\ref{fig3}) all have practically one and the same amplitude, $\sim$six times below that which is met usually at lightning discharges in thundercloud. In contrast to usual thunderstorm phenomena, these anomalous pulses seen in pre-earthquake period have not been accompanied with any infrasonic signal from the high-sensitive pressure variation sensor which has been operating simultaneously as a part of detector complex. (Remember also that there was a period of a quiet clear weather in the day of 1~May, 2011, without any sign of electric activity in atmosphere). As it is seen in the high-resolution incut to the top plot of figure~\ref{fig3}, the series of unidirectional positive pulses terminates 40~min before the first seismic shock, and reappears only around the moment of the most powerful 4.2~magnitude aftershock which has happened on 2:31~UT 2~May, 2011. Similar disturbances of the local electric field observed in a pre-earthquake period have been just reported previously (e.g. \cite{seismoele_morgunov,seismoele_greece1}).

If to remove all the moments of spike-like enhancement from the temporal record of electric field and to subject the remaining data to a low-pass filtering procedure the fine structure of field variation may be revealed which is shown in lower plot of the figure~\ref{fig3}. Transient irregularities in the field variation spectrum are seen on this plot which precede with a nearly one-hour accuracy the moments both of the main earthquake shock and a powerful aftershock which has happened $\sim$6~hours later.

In comparison of the figures~\ref{fig1}, \ref{fig2}, and \ref{fig3} one can see that the period of anomalous behavior of electric field begins simultaneously with the formerly discussed decrease in the gamma- and thermal neutron signal, and the most intensive electric activity falls on the moment of the deepest dip in the intensity of gamma-radiation and neutron flux, immediately before the main earthquake shock. The latter circumstance is another evidence against any random coincidence, and, contrarily, of a close correlation between these rarely met anomalies in behavior of the three heterogeneous geophysical characteristics.

\section{Conclusion}

Some characteristic influence of a 5.4~magnitude earthquake has been revealed in monitoring records of the intensity of environmental gamma-radiation, of the flux of low-energy neutrons, and of the local electric field at the complex detector installation of the Tien~Shan mountain station. Anomalous variations of these quite different geophysical parameters are in a good correlation with each other both during the $2-3$~days long period preceding the earthquake moment, and immediately before and in the time of seismic activity. The results presented here seem to be promising from the viewpoint of the forecast problem of powerful earthquakes though further investigations in this field still remain to be of a vital necessity.

\bibliographystyle{unsrt}
\bibliography{p}

\end{document}